\documentclass{aa}

\usepackage[]{color}
\definecolor{Red}{rgb}{0.9,0.17,0.31}
\definecolor{Green}{rgb}{0,0.5,0}
\definecolor{ashgrey}{rgb}{0.7, 0.75, 0.71}
\definecolor{bluebell}{rgb}{0.64, 0.64, 0.82}
\definecolor{darkorchid}{rgb}{0.6, 0.2, 0.8}
\definecolor{glaucous}{rgb}{0.38, 0.51, 0.71}
\definecolor{darkcerulean}{rgb}{0.03, 0.27, 0.49}
\definecolor{flame}{rgb}{0.89, 0.35, 0.13}
\definecolor{brightmaroon}{rgb}{0.76, 0.13, 0.28}
\definecolor{maroon}{rgb}{0.5, 0.0, 0.0}
\definecolor{carmine}{rgb}{0.59, 0.0, 0.09}
\definecolor{darkcyan}{rgb}{0.0, 0.55, 0.55}
\definecolor{electricultramarine}{rgb}{0.25, 0.0, 1.0}

\usepackage[utf8]{inputenc}
\usepackage[T1]{fontenc}
\usepackage{lmodern}
\usepackage{url}
\usepackage{booktabs}
\usepackage{amsfonts}
\usepackage{nicefrac}
\usepackage{microtype}
\usepackage{lipsum}
\usepackage[parfill]{parskip}
\usepackage{graphicx}
\usepackage[normalem]{ulem} 
\usepackage{tabularx}
\usepackage[colorlinks, citecolor=darkcerulean, urlcolor=glaucous,linkcolor=brightmaroon]{hyperref}
\usepackage{caption}
\usepackage{subcaption}

\usepackage{natbib}
\bibpunct{(}{)}{;}{a}{}{,}

\title{Predicting large-scale cosmological structure evolution with generative adversarial network-based autoencoders}

\author{Marion Ullmo\inst{1,2,3} \and Nabila Aghanim\inst{2} \and Aurélien Decelle\inst{3,4} \and Miguel Aragon-Calvo\inst{5}}

\institute{
\inst{1} IRFU, CEA, Université Paris-Saclay, Gif-sur-Yvette, France \\
\inst{2} Université Paris-Saclay, CNRS,  Institut d'Astrophysique Spatiale, B\^atiment 121 Campus Paris-Sud 91405, Orsay, France \\
\inst{3} Université Paris-Saclay, CNRS, TAU team INRIA Saclay, Laboratoire de recherche en informatique,
91190, Gif-sur-Yvette, France. \\
\inst{4} Departamento de Física Téorica, Universidad Complutense, 28040 Madrid, Spain \\
\inst{5} Instituto de Astronomía, UNAM, Apdo. Postal 106, Ensenada 22800, B.C., México
}

\authorrunning{A1}
\date{}

\abstract{Predicting the nonlinear evolution of cosmic structure from initial conditions is typically approached using Lagrangian, particle-based methods. These techniques excel in terms of tracking individual trajectories, but they might not be suitable for applications where point-based information is unavailable or impractical. In this work, we explore an alternative, field-based approach using Eulerian inputs. Specifically, we developed an autoencoder architecture based on a generative adversarial network (GAN) and trained it to evolve density fields drawn from dark matter N-body simulations. We tested this method on both 2D and 3D data. We find that while predictions on 2D density maps perform well based on density alone, accurate 3D predictions require the inclusion of associated velocity fields. Our results demonstrate the potential of field-based representations to model cosmic structure evolution, offering a complementary path to Lagrangian methods in contexts where field-level data is more accessible.}


\begin{document}

\maketitle

\section{Introduction} \label{Introduction}

Cosmological simulations allow us to confront existing theoretical models describing the Universe’s initial state, contents, and the physical processes governing its evolution with increasingly detailed observations of the cosmos, such as the data coming from James Webb Space Telescope \citep{gardner2006james} or the upcoming Euclid survey \citep{laureijs2011euclid}.
These simulations describe the evolution over time of matter on large scales, from N-body simulations modeling only the gravitational interaction of massive particles in expanding space (e.g., the Millenium simulation, \citealt{boylan2009resolving}) to more complex simulations (e.g., Illustris, \citealt{vogelsberger2014introducing}, IllustrisTNG, \citealt{nelson2019illustristng}, EAGLE, \citealt{crain2015eagle}) incorporating hydrodynamics and/or physical processes on relatively small scales modeling the evolution of baryonic matter.

When considering the N-body problem, we must keep in mind that there is no known closed-form expression that can be used to directly calculate the positions and velocities of the particles at any given time. Indeed, the gravitational force between each pair of DM particles in simulations depends on their positions and masses, while the resulting motion is influenced by the collective interactions among all the particles, entailing a highly nonlinear evolution of the system over time. Current numerical methods tend to approach this problem by starting with initial conditions of particle positions and velocities and iteratively resolving the equations of motion in time steps that are small enough that a linear evolution sufficiently approximates the true motion of particles. 
While gravitational interactions between particles scale naively as $\mathcal{O}(N^2)$, many efficient algorithms, such as particle-mesh (PM - $\mathcal{O}(N+M\log M)$, \citealt{efstathiou1985numerical}), tree methods ($\mathcal{O}(N \log N)$, \citealt{barnes1986hierarchical}), and hybrid approaches such as $\textrm{P}^3\textrm{M}$ ($ \sim\mathcal{O}(N \log N)$), have been developed to reduce this cost substantially. Nevertheless, high-resolution simulations with large particle counts remain computationally intensive, particularly when one aims to capture both the large-scale structure and small-scale dynamics simultaneously. This imposes practical limits on the resolution, scale, or number of realizations that can be feasibly simulated.
Although faster fully analytical approaches \citep{shandarin1989large, kitaura2013cosmological} and semi-analytical simulations that combine traditional simulation methods and analytical approximations \citep{monaco2002pinocchio,tassev2013solving}, relying on first- or second-order perturbation theory have been proposed, they cannot address the highly nonlinear stages of structure formation.
 
\begin{figure*}
  \centering
  \includegraphics[width=.8\textwidth]{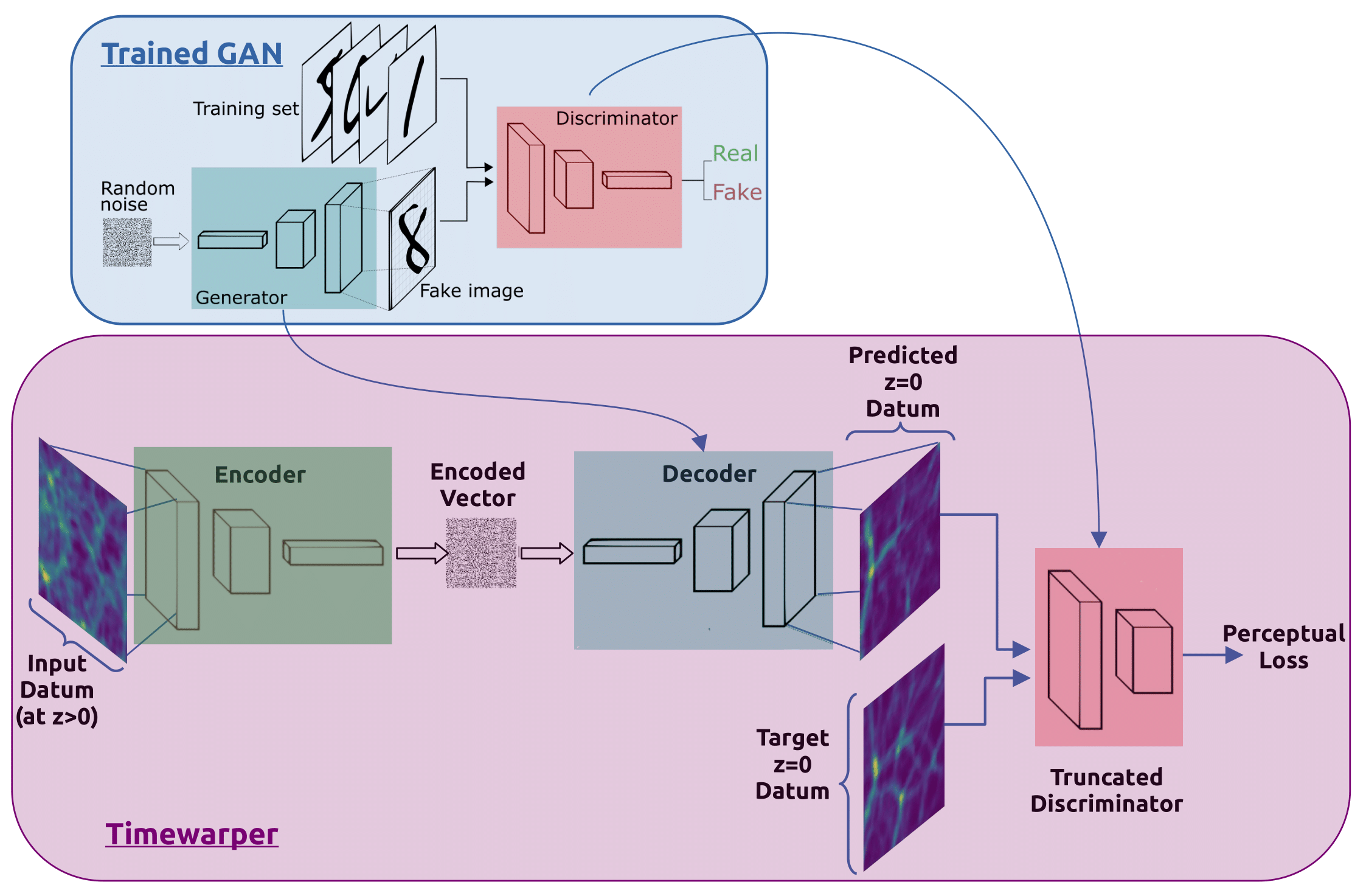}
  \caption{Architecture of the timewarper. A trained GAN's generator is used as a readily built decoder. Only the encoder's weights are changed during training. The same GAN's truncated discriminator is used to compute a perceptual loss (see Eq. \ref{eq:AEloss}).}
  \label{fig:TWarchi}
  
\end{figure*}

More recently, machine learning (ML) approaches have proved efficient in completing tasks for which prescriptive analytical approaches were either slow, overly approximative, or nonexistent. Examples in astronomy include classification, such as galaxy types \citep{shamir2009automatic,freed2013application,biswas2018classification} or cosmic web structure types \citep{aragon2019classifying}, as well as redshift estimation \citep{henghes2021benchmarking, rastegarnia2022deep, d2018photometric, henghes2022deep}, detection of various objects of interest \citep{rezaei2022decoras, vafaei2019deepsource, jia2023deep, gonzalez2018galaxy}, anomalies \citep{reyes2020transformation, dere2021anomaly, villar2021deep}, or physical effects such as gravitational lensing\citep{rezaei2022machine, wilde2022detecting} or Sunyaev-Zel'dovich (SZ) effects\citep{tanimura2022convolutional, bonjean2020deep}, deblending sources \citep{burke2019deblending, hausen2022partial, hansen2022scalable}, determining baryonic matter properties given dark matter distribution \citep{jo2019machine, wu2023learning, chittenden2023modelling, bonjean2019star}, simulation augmentation \citep{sweere2022deep, kodi2020super, agarwal2018painting}, and others.
More specifically, when considering the time evolution of complex physical systems, several neural network-based experiments \citep{feng2023predicting, wiewel2019latent, liu2023dynamics, humbird2018predicting, sanchez2020learning} have demonstrated promising results in terms of emulating simulations.

Concerning cosmological simulations in particular, recent years have seen the emergence of a range of machine learning (ML) approaches designed to accelerate the generation of N-body simulations, offering  speed-ups of several orders of magnitude over traditional numerical methods. A particularly successful class of these models is rooted in a Lagrangian framework, first introduced with the seminal $\textrm{D}^3\textrm{M}$ model \citep{he2019learning} and subsequently refined to impressive levels of performance in later works \citep{de2020fast,jamieson2023field,jamieson2024field,prost2025reseau}. These models typically take the initial positions of particles, together with a linear approximation of their evolution (e.g., the Zel’dovich approximation) and learn to predict the final nonlinear displacement by estimating the residual between this approximation and the true evolution.
Beyond the direct emulation of structure formation, this framework has also been adapted for broader use cases, such as translating simulations across different cosmological parameters or initial conditions \citep{saadeh2024field,giusarma2023learning}. These methods have proven highly effective in recovering the small-scale clustering and halo structures characteristic of dark matter evolution.
Finally, hybrid strategies that combine ML approximations with partial N-body integration, such as the COCA model \citep{bartlett2024comoving}, have further demonstrated the potential of ML-assisted simulation pipelines, achieving improvements in both speed and accuracy  in cosmological modeling.

While these approaches are optimized for the specific task of accelerating particle-based simulations, complementary methods can be explored through a Eulerian perspective, namely, by treating the matter distribution as a continuous field rather than a set of discrete particles. 
Although less optimal than Lagrangian approaches in this context, focusing on Eulerian fields offers a simplified, but generalizable framework.
This can help inform other ML-based inference challenges where only Eulerian data are available; for instance, it naturally aligns with the structure of specific types of observational data, such as weak lensing maps, 21cm intensity maps, or cosmic microwave background lensing fields, all of which are inherently gridded or projected fields. Similarly, many hydrodynamical simulations represent gas and baryonic components in Eulerian form, making field-based ML architectures well-suited for cross-domain applications \citep{hsu2024reconstructing,luo2025galaxy,gondhalekar2025emulation,hiegel2023reionisation, conceiccao2024predicting, conceiccao2024fast, li2022high}.

Moreover, the two paradigms are complementary in their scaling limitations. Lagrangian methods are typically constrained by the number of particles that can be simulated or stored, while Eulerian approaches are limited by voxel resolution and binning artifacts. There exists an intermediate regime, where extremely fine particle simulations are prohibitively expensive for particle-based ML, yet still accessible to Eulerian models through lower-resolution gridding. In such cases, the field-based approach enables analysis of structure formation at finer graining and with potentially better memory efficiency.

In this context, we propose a proof-of-concept model aimed at predicting the time evolution of cosmic density fields in Eulerian space. Rather than offering a drop-in replacement for Lagrangian-based emulators, this work explores the feasibility of using generic continuous fields as inputs, with the long-term goal of developing architectures whose application can be generalized to a range of cosmological and astrophysical datasets.

To this end, we have introduced an autoencoder (AE) architecture with an unconventional twist: it incorporates generative adversarial network (GAN) components into its architecture \citep{ullmo2021encoding}. In the following, we refer to this AE as timewarper (TW). This design encourages the model to balance two objectives: an accurate reconstruction of individual structures and the preservation of the overall statistical properties of the density field. While this hybrid approach currently exhibits a tradeoff between spatial precision and statistical fidelity compared to standard autoencoders, we argue that such tradeoffs are an essential aspect of the design space and exploring them may ultimately lead to more robust emulation strategies.

In Sect. \ref{sec:setup}, we outline the setup of our experiment, wherein we describe the architecture and training of our TWs, the data used for training, and the metrics by which we measure the quality of our results. 
In Sect. \ref{bl_results}, we begin by showing results for a TW trained to predict simply from an input density map; we refer to it as a "baseline TW."
In Sect. \ref{vel_results}, we introduce a TW trained with additional input information in the form of the density field's associated velocity field, which shows significant improvement on the baseline approach. We label it here as "velocities TW."
In Sect. \ref{sec:discussion}, we interpret our results and discuss other possible optimization methods.
Finally, we present our conclusions in Sect. \ref{sec:conclusion}.

\section{Setup}\label{sec:setup}

Our goal in this work is to create a network capable of forecasting the evolution of a simulation-derived data set (a discrete 2D or 3D density map generated from a 2D or 3D N-body simulation) from previous instances (when $z>0$) to the current time ($z=0$). Thus, we have built a TW that takes a datum in the form of a simulated density field at a given fixed redshift (i.e., $z=0,1,2$ or $3$) as input and is tasked with recovering its corresponding simulation at $z=0$, by minimizing the distance (see eq.\ref{eq:AEloss}) between its output and target datum.

While this work focuses on inputs at $z=3, 2, 1$, and $0$, incorporating earlier snapshots, such as the initial conditions ($z=99$), remains a potential avenue for future investigation. In a second step, we provide the TW an associated velocity field, in an effort to improve the AE's predictions. We  present the networks, data, and metrics used for our work in brief below.

\begin{figure*}
  \centering
  \includegraphics[width=.8\textwidth]{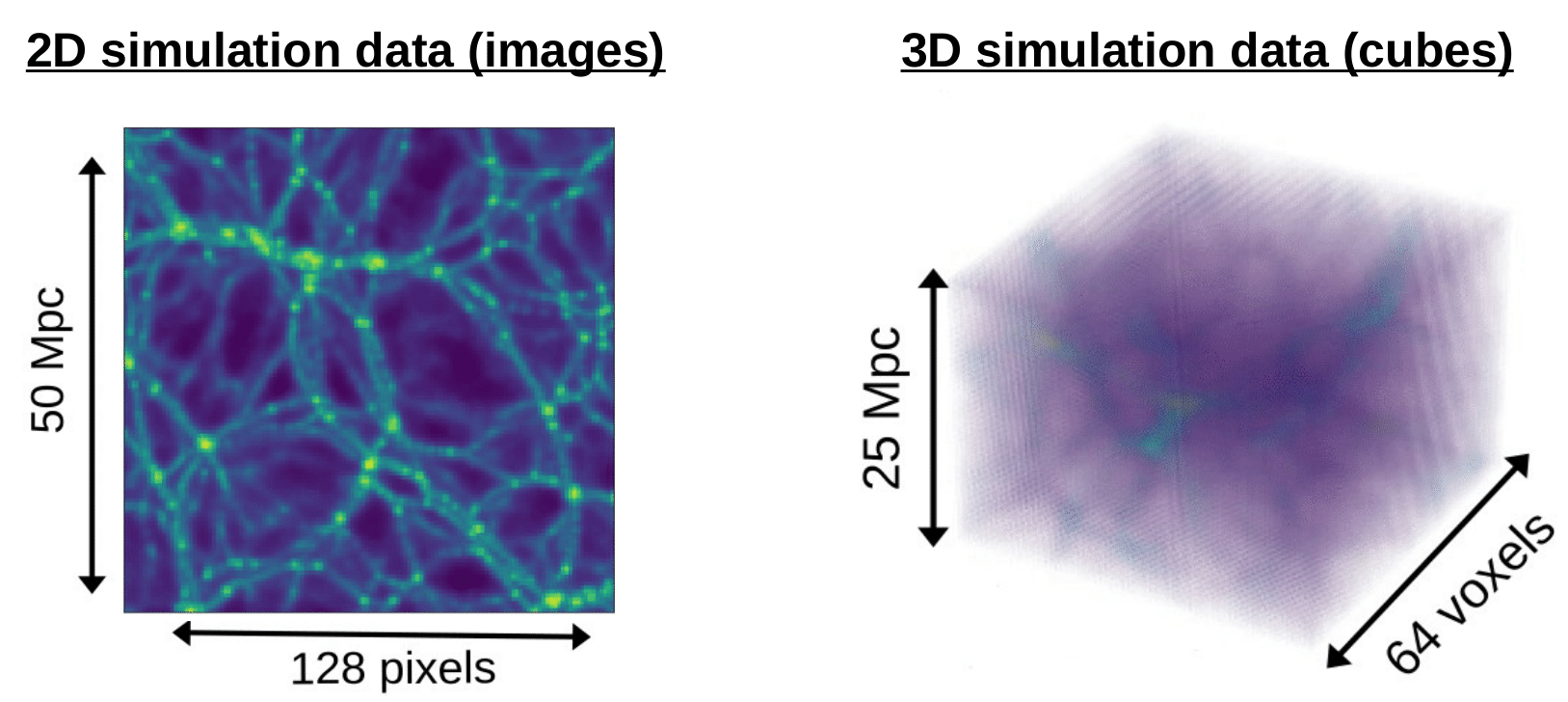}
  \caption{Example of simulation-issued data: a 2D image from a 2D simulation({\it left}) and a 3D cube ({\it right}) from a 3D simulation.
  These data are built by dividing an N-body simulation snapshot into pixels (or voxels) and counting the particles within each, creating a discrete density map. The map is then smoothed with a Gaussian filter and log-transformed, allowing cosmic structure to stand out starkly. This process creates data that are compatible with convolutional neural networks, which specialize in feature detection. }
  \label{fig:pixelpdf}
  
\end{figure*}

\subsection{Networks}
\subsubsection{GANs in a nutshell}

Overall, GANs are notorious for their ability to emulate data from a training set, typically in the form of images. They consist of a pair of deep neural networks that are trained simultaneously in an adversarial manner (see top diagram in Fig.\ref{fig:TWarchi}). The first, a generator, intakes a vector with randomly distributed values and is trained to output realistic data by "fooling" the second, a discriminator, which is tasked with separating genuine training data from data generated by the generator. Throughout training they become gradually better at their respective tasks, simultaneously increasing the difficulty of the other's task. Once trained the generator is able to output a training set-like datum from a low-dimension input vector, while the discriminator is capable of distinguishing high-level features in data to suit its task. We made use of these properties to build our AEs.

\subsubsection{Timewarpers}

Our work relies on the use of AEs based on these convolutional GANs (the TW, displayed in the bottom panel of Fig.\ref{fig:TWarchi}). The AEs consist of an encoder which intakes a datum at a given redshift $z\geq0$ and encodes it into a vector within a low-parameter latent space with greater semantic significance. This encoded vector is then fed into a decoder tasked with recovering the original datum at redshift $z=0$, by minimizing the loss described below (Eq. \ref{eq:AEloss}).

We aimed to constrain the decoder to output data that would be statistically consistent with true $z=0$ data. To this effect, we relied on a GAN that had been previously trained to generate simulation-like data at $z=0$. First, we used its generator (designed to output simulation-like data at $z=0$) as the AE's decoder and locked its weights during training. Second, we used its truncated discriminator as part of the AE's loss\footnote{See original article \citep{ullmo2021encoding}},

\begin{equation}\label{eq:AEloss}
    L_{AE}=\Delta(x,\tilde{x})
,\end{equation}

where $x$ is the output datum (i.e., the AE's prediction of input datum at $z=0$), $\tilde{x}$ is the ground truth (true input datum at $z=0$) and $\Delta$ is the $\ell_2$ difference in the truncated discriminator's latent space.

This type of loss, termed perceptual loss \citep{johnson2016perceptual}, integrates a discriminator's ability to emphasize the features and patterns of data rather than relying solely on a basic pixel-by-pixel comparison. The TW architectures and parameters (Table \ref{tab:temps}) are derived from previously trained 2D and 3D GANs, which have been adapted, in turn, from a publicly available DCGAN implementation\footnote{Tiago Freitas, \url{https://github.com/tensorfreitas/DCGAN-for-Bird-Generation}}. Notably, the latent dimension was inherited from the original 2D model and doubled for the 3D version to account for the higher input dimensionality. This scaling was empirically found to provide stable training and satisfactory predictive accuracy; larger latent spaces were summarily tested but did not appear to noticeably improve performance. Nonetheless, further exploration might be worthwhile in future work.

Notably, the model does not have any mass conservation constraints at present. We did not apply a mass normalization to the output density data, as a simple post-processing normalization would erroneously shift all of the amplitudes of the predicted structures. One possible improvement would be to introduce a normalization constraint during training, either by adding a dedicated loss term or by normalizing the data immediately before computing the existing loss.

\begin{table}[h!]
\caption{Architectures of the 2D and 3D TW.}
 \begin{subtable}[h]{0.45\textwidth}
    \caption{2D TW encoder (E) and decoder (D) specifications.}
    \centering
    \begin{tabular}{|c|c|} 
    \hline
    Input shape & $128\times128$ \\
    \hline
    Filter sizes & $\{5, 5, 5, 5, 5\}$ \\
    \hline
    $n_{filter} (E)$ & $\{32, 64, 128, 256, 512\}$ \\
    \hline
    $n_{filter} (D)$  & $\{256, 128, 64, 32, 1\}$ \\
    \hline
    Strides:  &  $\{2, 2, 2, 2, 2\} $ \\
    \hline
    Layer activation & Leaky ReLU (E), ReLU (D) \\
    \hline
    Final activation & None(E), Tanh (D) \\
    \hline
    Latent dimension & $100$ \\
    \hline
    \end{tabular}
    
    \label{tab:2D_archi}
 \end{subtable}
 \hfill
 \begin{subtable}[h]{0.45\textwidth}
    \caption{3D TW encoder (E) and decoder (D) specifications.}
    \centering
    \begin{tabular}{|c|c|} 
    \hline
    Input shape & $64\times64\times64$ \\
    \hline
    Filter sizes & $\{4, 4, 4, 4\}$ \\
    \hline
    $n_{filter} (E)$ & $\{32, 64, 128, 256\}$ \\
    \hline
    $n_{filter} (D)$  & $\{128, 64, 32, 1\}$ \\
    \hline
    Strides:  &  $\{2, 2, 2, 2\} $ \\
    \hline
    Layer activation & Leaky ReLU (E), ReLU (D) \\
    \hline
    Final activation & None(E), Tanh (D) \\
    \hline
    Latent dimension & $200$ \\
    \hline
    \end{tabular}
    
    \label{tab:3D_archi}
 \end{subtable}
 \tablefoot{All models are trained using the {\it Adam} optimizer with the same parameters as above ($lr=0.0002, \beta_1=0.5$).}
 \label{tab:temps}
\end{table}

\subsection{Data}
\subsubsection{Simulations}
Our networks were trained on data built from both 2D and 3D N-body simulations. The 2D data (Fig. \ref{fig:pixelpdf}, left) provide simpler conditions (i.e., a lower parameter problem in terms of the particle degree of freedom, datum size, and network size), leading to more optimal results for a more ideal proof of concept and a point of comparison for 3D results. 
The training set is built from $1000$ snapshots from nbody2D\footnote{credit: Johannes Hidding\\ \url{https://zenodo.org/record/4158731\#.X5_ITJwo-Ch}} simulations, a set of 2D particle-mesh N-body simulations of side $100\textrm{Mpc}/h$ with $512^2$ particles using the standard $\Lambda$CDM cosmology. They were run and saved for redshifts $z= 0, 1, 2,$ and $3$.

The training set for the 3D data of interest to our study (Fig. \ref{fig:pixelpdf}: right panel) was built from a single snapshot of a GADGET2 simulation \citep{GADGET1,GADGET2} of a side of $100 \textrm{Mpc}/h$ with $512^3$ particles. We assumed a standard $\Lambda$CDM cosmology, with cosmological parameters $\Omega_m=0.32$, $\Omega_\Lambda=0.69$, $\sigma_8=0.83$, $n_s=0.96$, and $H_0=0.68,$ from Planck 2018 \citep{aghanim2018planck} and saved at redshifts $z= 0, 1, 2,$ and $3$.

\begin{figure*}
    \centering
    \includegraphics[width=\textwidth]{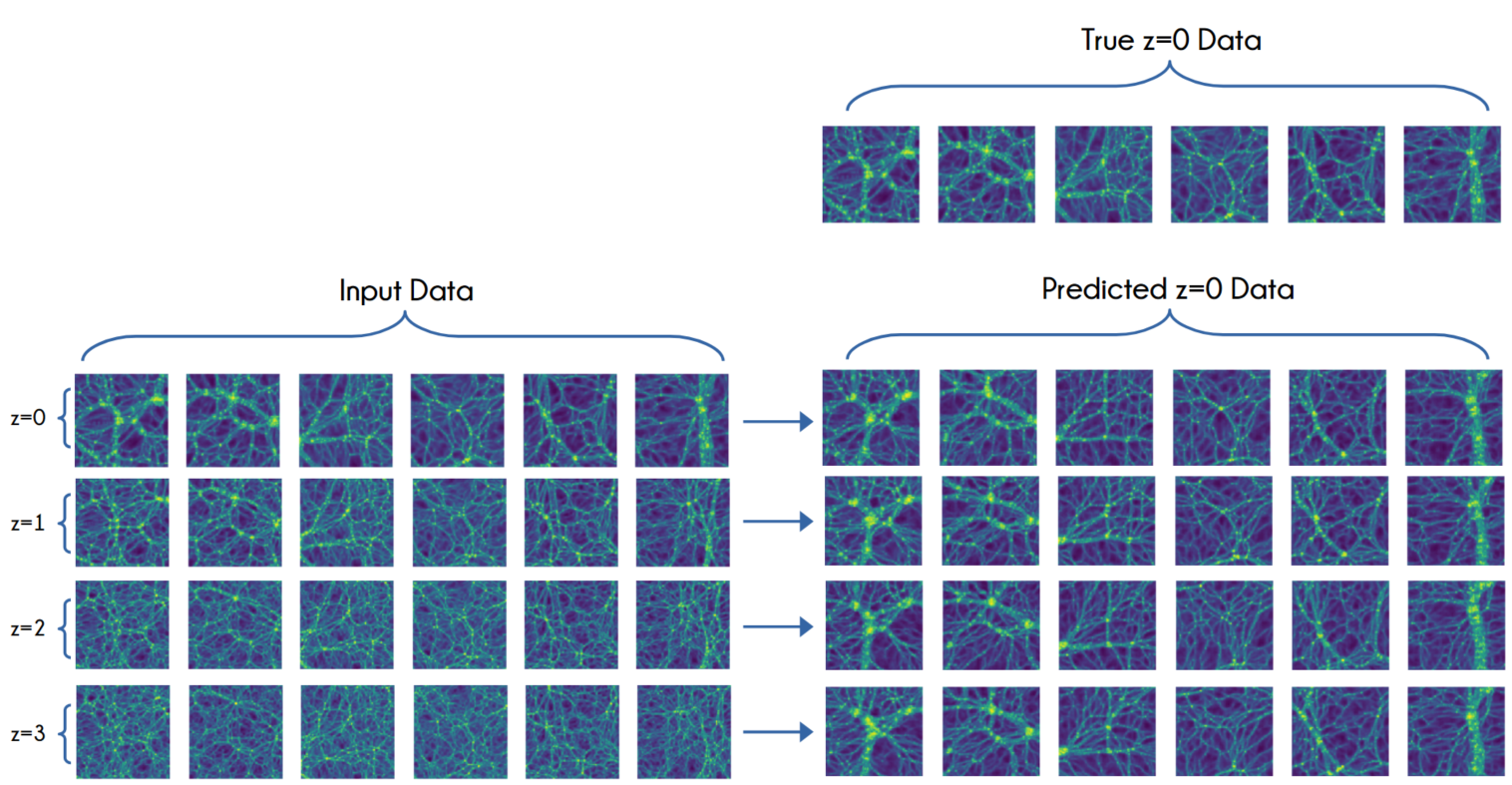}
    \caption{Six images from the 2D simulations at various redshifts ({\it left}), and their equivalent predictions of redshift $z=0$ ({\it right}) inferred by the {\it baseline TW}. The true $z=0$ simulation images are shown above the predicted images ({\it upper right}) for comparison.}
    \label{fig:2D_TW_baseline_ims}
\end{figure*}

\subsubsection{Training data}

The individual data were built in the following way: each (2D or 3D) snapshot is first made into a discrete density map ($256\times256$ or $256\times256\times256$, respectively).
For the 3D data, these were constructed following a crude nearest grid point (NGP) procedure with subsequent smoothing using a Gaussian filter with a standard deviation of 3 pixels. This smoothing reduces sharp discontinuities in the raw density field that arise from coarse binning noise.
The 2D data were built using a Delaunay tessellation field estimator (DTFE, \cite{aragoncalvo2020smooth}).
They were subsequently log-transformed for compatibility with the neural network. The choice for the 3D data was made for simplicity, but further work would likely benefit from using more accurate interpolation schemes such as the 2D data's DTFE, cloud-in-cell, triangular-shaped cloud \citep{hockney2021computer}, or simplex-in-cell \citep{Abel2012Tracing,Hahn2012New}.

From these data, smaller sub-arrays (of side $50 \textrm{Mpc}/h$ and $128$ pixels for 2D and $25 \textrm{Mpc}/h$ and $64$ voxels for 3D) were extracted to make up the training sets. This gives us respectively $~5.10^8$ and $~8.10^8$ possible sub-arrays.
Given this high number of possibilities and the redundancy among them, we did not reason in terms of epochs for the training time (where one epoch corresponds to the network processing the entire training set). 
Instead, we quantified the training time in terms of gradient updates performed on data batches (of size $200$ in our case), with each training batch randomly selected from all possible subarrays.

In a second part of our work, we additionally provided the density fields' associated velocity fields as input for the 3D case.
To build the velocity field, we followed a similar approach to our consideration of the density field. Dividing the snapshot space into $3\times256\times256\times256$ voxels, we computed three 3D averaged velocity fields for each direction (x, y, and z), by summing the velocities of the particles in each voxel and dividing the sum by the number of particles.
Upon inspecting the resulting cubes, we find that cosmic structures are visually apparent without need for log-transformation. Thus, we can simply apply a normalization of voxel values, 

\begin{equation}
    v'=v/N
.\end{equation}

Here, $v'$ is the new velocity and N is fixed such that  $|v'|_{max}\lesssim1$.
Next, we can apply the same smoothing as for the 3D density to obtain our final three $256\times 256\times 256$ velocity fields for each (x, y, and z) direction.
By combining the density field with the velocity fields  constructed in this way, we can construct an array of size $256\times 256\times 256\times 4   $ and use it to extract smaller arrays of size $64\times 64\times 64\times 4$.

For the 3D case, the training, validation, and test sets were constructed from three distinct simulations. The validation set consists of 200 randomly selected sub-cubes from the validation simulation, while the test set comprises of 64 sub-cubes forming the test simulation. In the results section, we reconstructed the full cube by assembling the 64 predicted sub-cubes and compared the result with the corresponding full cube at the target redshift.

For the 2D case, we used 1000 simulations in total: 800 for training, 100 for validation, and 100 for testing. Each simulation is divided into four sub-arrays. Consequently, the test set contains 400 predicted subarrays, which were reassembled into 100 full-size arrays for evaluation.

\begin{figure*}
    \includegraphics[width=\textwidth]{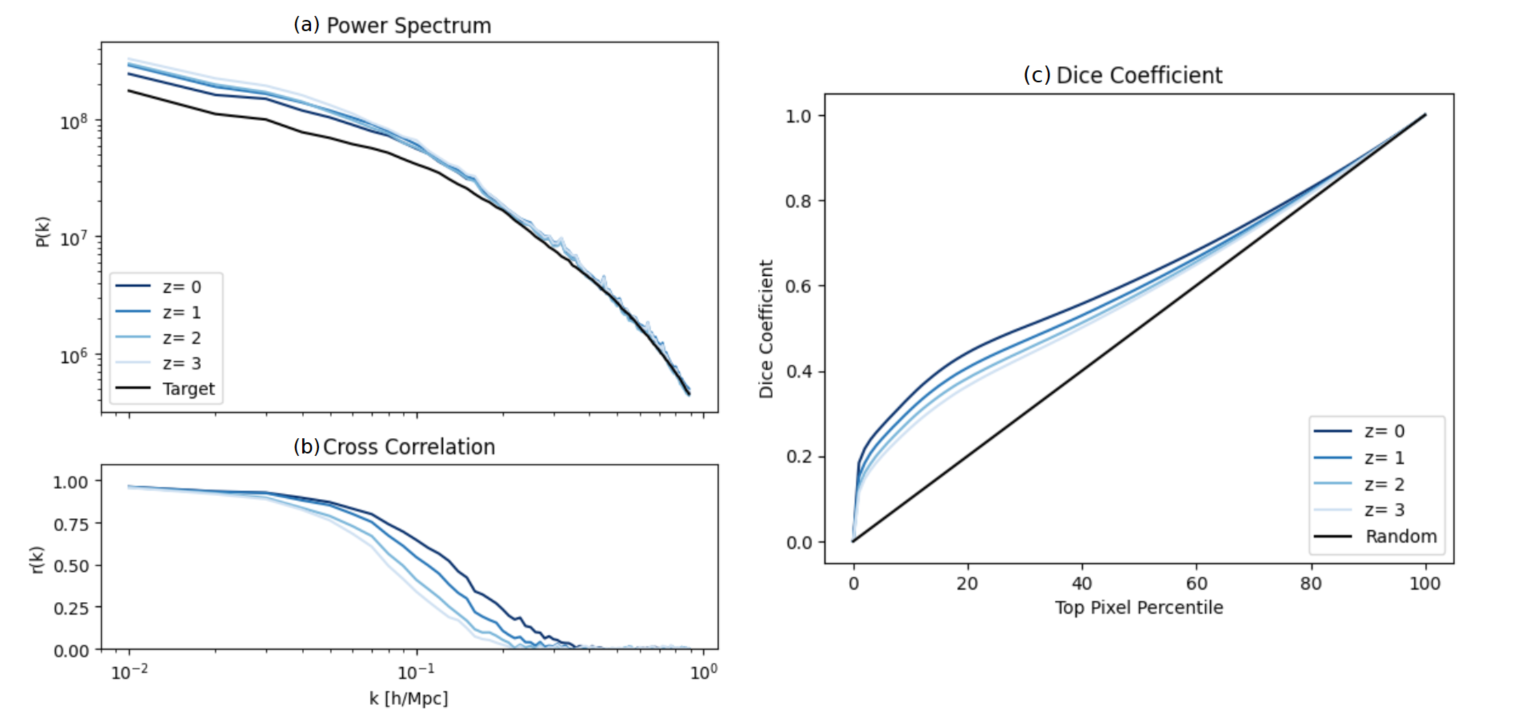}
    \caption{Spectra, cross-correlation coefficient and Dice coefficient for baseline timewarper on 2D data. {\it (a)} Average power spectra from predictions from input redshifts z = 3 → 0 (blue scale), and average target spectrum at z = 0 (black).
{\it (b)} Corresponding average cross-correlation coefficient between prediction and target over the same redshift range.
{\it (c)} Average Dice coefficient between prediction and target (same blue color scale as in {\it a} and {\it b}).}
    \label{fig:2D_TW_baseline_ps_ovl}
\end{figure*}

\subsection{Metrics}

To quantify the efficiency of the TW’s predictions, we relied on three metrics: the power spectrum and the cross-correlation coefficient, which assess the spatial correspondence between predicted and target data; along with the Dice coefficient (see \cite{ullmo2021encoding} for details), which measures the accurate recovery of dense structures in the predictions. While the power spectrum provides a global view of the statistical consistency between predicted and target data, the cross-correlation and Dice coefficients additionally evaluate the pairwise agreement of individual targets and predictions (in other words, the TW's accuracy). The cross-spectrum of a data pair $A$ and $B$ for a frequency, $k$, is given by

\begin{equation}
P_{AB}(k)=\langle A_{\bf k} B^*_{\bf k} \rangle_{{\bf k}, |{\bf k}|=k}, 
\end{equation}

where $A_{\bf k}$ and $B_{\bf k}$ are the pair's discrete Fourier transform elements. 

From this, we get the power spectrum of datum $A$: $P_{AA}(k).$

Since the power spectrum's distribution is approximately uniform in logarithmic space, when averaging over the test set we use the geometric mean, $\exp\!\left(\langle \ln P \rangle\right)$, as a more representative central value.
Variability measures (e.g., the geometric standard deviation) were considered in our study, but omitted from the figures for clarity.

Additionally, we can compute the cross-correlation coefficient for pair $A$ and $B,$ 

\begin{equation}
r(k) = \frac{P_{AB}(k)}{\sqrt{P_{AA}(k),P_{BB}(k)}},
\end{equation}

The Dice coefficient for a data pair $A$ and $B$ is expressed as

\begin{equation}
O_{AB}(t)=\frac{N_{A\cap B} (t)}{N_{A\cup B} (t)} \label{eq:simple_ovl}    
,\end{equation}

where $N_{A\cap B}(t)$ is the number of pixels or voxels whose value is above $t$ for both $A$ and $B$ and $N_{A\cup B}(t)$ is the number of pixels or voxels whose value is above the threshold, $t$, in either $A$ or $B$ at a given position in an image or cube.
We computed this across a fixed set of thresholds corresponding to percentiles of the total pixel value distribution.
This metric quantifies the degree of pixel-wise overlap between the predicted and true high-density regions.
Assuming the pixel value distributions of the predicted and target maps are broadly similar, the Dice coefficient of an unrelated pair of data at a given threshold is expected to approximate the fraction of pixels above that percentile. In contrast, a perfect reconstruction would yield a Dice coefficient of $1.0$ across all percentiles. This is an ideal, but challenging scenario given the complexity of exact pixel-level alignment in continuous fields.

\begin{figure*}
    \centering
    \includegraphics[width=\textwidth]{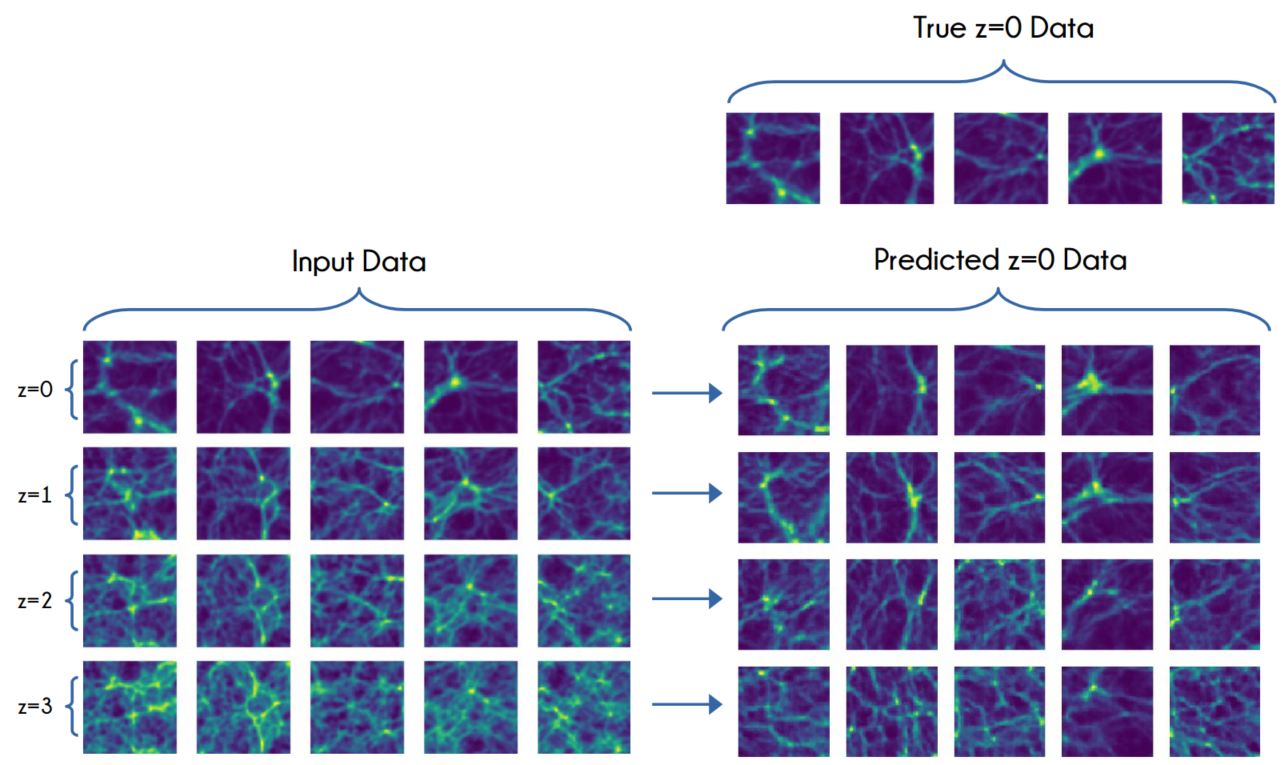}
    \caption{Five images from the 3D simulations at various redshifts ({\it left}) and their equivalent predictions of redshift $z=0$ ({\it right}) as inferred by the {\it baseline TW}. The true $z=0$ simulation images are shown above the predicted images ({\it top-right}) for comparison.}
    \label{fig:3D_TW_baseline_ims}
\end{figure*}

\begin{figure*}
    \centering
    \includegraphics[width=\textwidth]{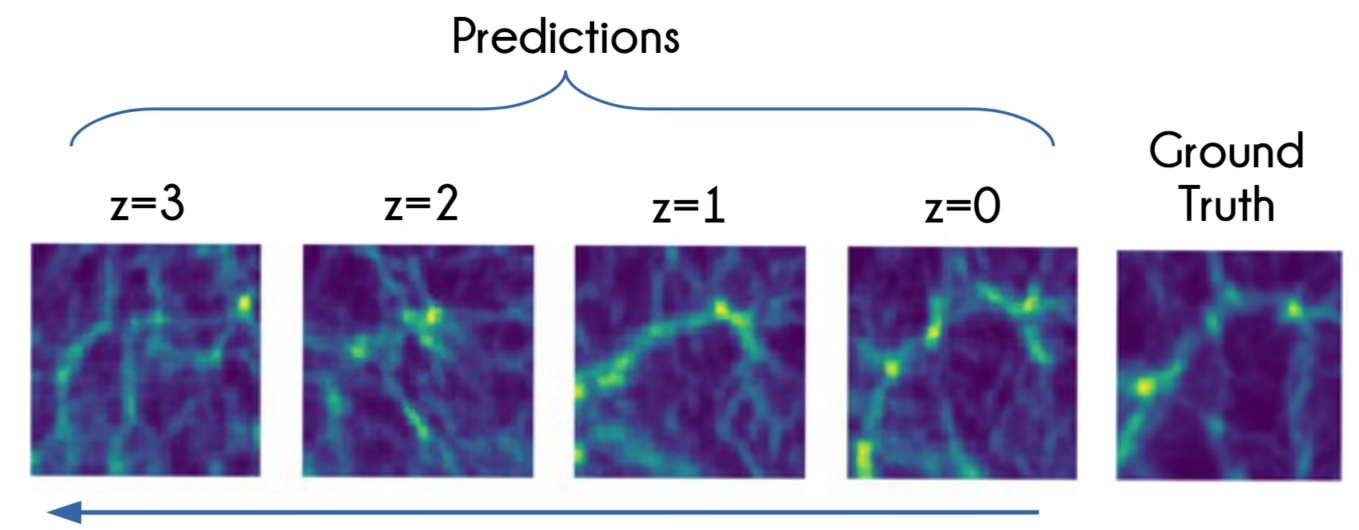}
    \caption{Closeup of predictions of a single datum of the 3D simulations for inputs $z=0, 1, 2$ and $3$, using the baseline TW. We can see that as $z$ grows, predictions are increasingly imprecise, favoring more homogeneous, underdense structures, compared to the ground truth's dense concentrated structures.}
    \label{fig:3D_TW_baseline_ims_closeup}
\end{figure*}

\newpage
\section{Results} \label{sec:results}

\subsection{Baseline TW results}\label{bl_results}

We input a datum in the form of a density field at redshift $z\geq0$ and trained the TW to output the corresponding density field at $z=0$. Here, “corresponding” refers to the same comoving region within snapshots at the input and target redshifts ($z_{ini}\longrightarrow z_{fin}$). The network was optimized by minimizing the perceptual loss (see Eq.~\ref{eq:AEloss}). We carried out this procedure for both 2D and 3D data.

\begin{figure*}
    \centering
    \includegraphics[width=\textwidth]{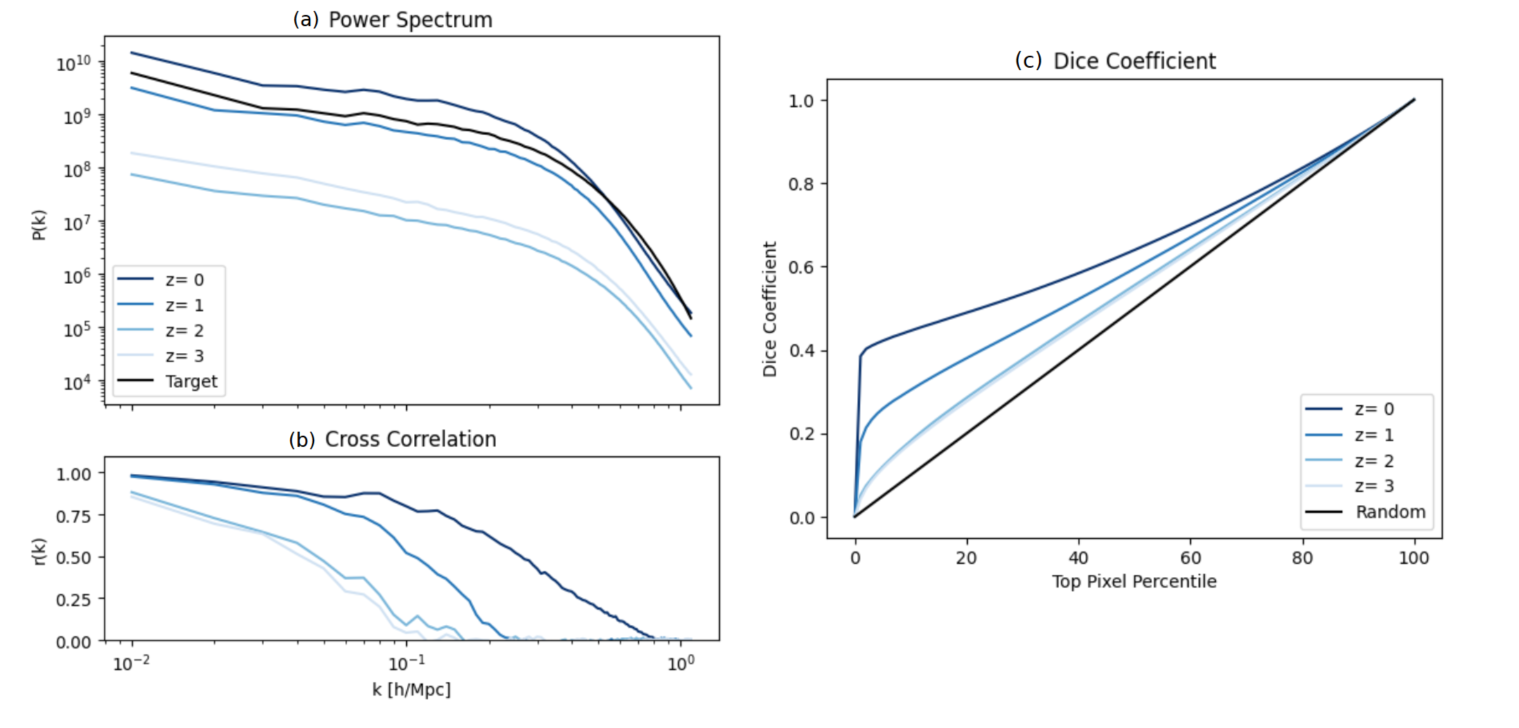}
    \caption{Spectra, cross-correlation coefficient and Dice coefficient for baseline timewarper on 3D data. {\it (a)} Power spectra from predictions from input redshifts z = 3 → 0 (blue scale), and target spectrum at z = 0 (black). {\it (b)} Corresponding cross-correlation coefficient between prediction and target over the same redshift range. {\it (c)} Dice coefficient between prediction and target (same blue color scale as in {\it a} and {\it b}).}
    \label{fig:3D_TW_baseline_ps_ovl}
\end{figure*}

\subsubsection{2D images}

We first focus on the outcome of the TW  trained to recover density fields at redshift $z=0$ for the set of 2D simulation images, from input density fields at redshifts varying from $z=0$ to $z=3$ by steps of $\Delta z=1$. We note that the "predictions" from $z=0$ to $z=0$ consist of simple replicative autoencoding, which define the limitations of our setup, the discrepancy between input and output in this case being strictly due to encoding loss, rather than inaccurate prediction.

We compared our results on predictions from higher redshifts to this $z=0$ reference. The task is expected to be harder the higher the input $z$; indeed, given that the development of structure in matter is a highly nonlinear process, we can expect that the farther away in time the target is from the output, the farther the density field will steer from an easily approximated linear evolution. For all $z>0 \longrightarrow z=0$ input-output pair, a distinct TW is individually run for 15k gradient updates (no significant improvement observed for longer training) and the best weights are recovered by finding the minima of validation losses computed every 1000 update for each input $z$ (0, 1, 2, and 3), respectively at 10, 15, 10, and 12k updates.

We illustrate the baseline TW's performance in recovering $z=0$ from different redshifts with a set of six simulation images taken at random from the test set (Fig. \ref{fig:2D_TW_baseline_ims}). These data are taken at various redshifts (left block) and used as input for the trained TWs to predict their $z=0$ equivalent (upper right). The predicted data are shown in the right block.

We find that regardless of input redshift, the TWs are successful in recovering $z=0$. The larger and denser structures are consistently well recovered while finer details exhibit more variability. Initially, it is difficult to discern any perceptible changes in performance based on varying input redshifts; indeed, structures appear to be reliably predicted even when inferring from higher redshifts. However, upon closer examination, we note that as the input redshift increases, there is a gradual loss of detail in the predicted structures. This loss  is manifested as the merging of certain large structures, shifts, or disappearance of finer structures, as well as alterations in the positions of overdense regions.

Inspecting the average power spectrum (PS) and cross-correlation coefficient $r(k)$ of the predicted data (Fig.~\ref{fig:2D_TW_baseline_ps_ovl}, left), we found that the TWs recover similar statistics, regardless of the input redshift. An upward shift of the PS, together with $r(k) \approx 1$ at low $k$, suggests that the predictions are coherent with the ground truth on large scales but slightly overestimate the total mass; for all metrics, standard deviations over the test set (left out of the figure for clarity) are relatively small compared to the measured averages for both target and predictions. The correlation begins to decline around $k \simeq 3\times10^{-2}\,h\,\mathrm{Mpc}^{-1}$ and reaches $r(k)=0$ between $k \simeq 2\!\rightarrow\!4\times10^{-1}\,h\,\mathrm{Mpc}^{-1}$ for the $z=3\!\rightarrow\!0$ case. Over this range, the PS curves overlap well, indicating that the model reproduces small-scale amplitudes but misplaces structures, leading to phase decorrelation. Since structures at these scales arise from highly nonlinear processes, this suggests that the model, while accurately capturing large-scale linear evolution, struggles more with the nonlinear regime. However, as some decorrelation at high $k$ is also visible in the simple $z=0\!\rightarrow\!0$ autoencoding case, part of this limitation likely stems from the encoder itself, with the latent space not perfectly representing all possible configurations at these scales.

\begin{figure*}
    \centering
    \includegraphics[width=\textwidth]{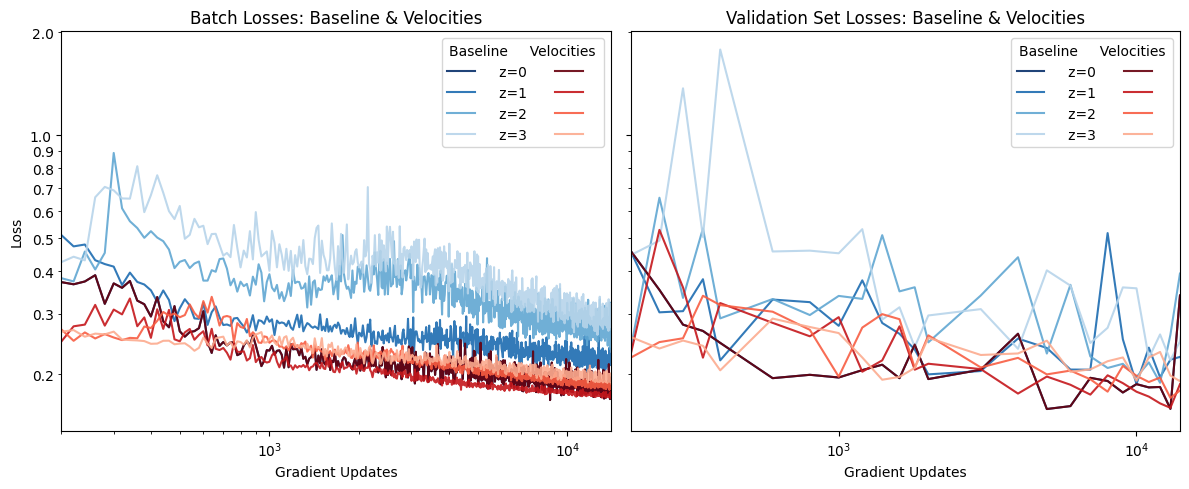}
    \caption{Batch (left) and validation set (right) losses for the velocities TW (red) and baseline TW (blue) at various input redshifts. Both panels share the same y-axis for easier comparison.}
    \label{fig:3D_TW_vel_losses}
\end{figure*}

The Dice coefficient (Fig. \ref{fig:2D_TW_baseline_ps_ovl}, right) shows that predictions across all redshifts align with the base $z=0 \longrightarrow 0$ autoencoding, but each increment in input $z$ yields a slightly lower coefficient at all pixel thresholds. This reflects the growing challenge of prediction as $z$ increases. From this first test on 2D data, we can conclude that (in this simple case at least) the TW can effectively predict structure evolution over the tested time spans. Its accuracy is seen to improve as the input redshift approaches the target.

\subsubsection{3D cubes}

We now focus on the outcome of the TW  trained to recover density fields at redshift $z=0$ for the set of 3D simulation cubes, from input density fields at redshifts varying from $z=0$ to $z=3$ by steps of $\delta z=1$.
The baseline TW is run for 30k gradient updates, and the best weights are recovered for each input $z=(0,1,2,3)$, respectively at 30, 30, 20, and 30k updates. Indeed, beyond a certain threshold the progression of loss appears to exhibit a linear pattern when plotted on a log-log scale (Fig. \ref{eq:AEloss}). We halted the training at 30k updates due to time constraints.

We observed five simulation cubes taken at random from the test set (Fig. \ref{fig:3D_TW_baseline_ims}). These data are taken at various redshifts ({\it left block}) and used as input for the trained TWs to predict their $z=0$ equivalent ({\it upper right}). The predicted data are shown in the right block. Here, we find that, contrary to the 2D case, the predicted data become noticeably more inaccurate (see Fig. \ref{fig:3D_TW_baseline_ims_closeup}); that is to say, we start to see the high-density contrast disappearing, while spurious low-density structures arise with the increase of the input redshift. The data predicted from $z=2$ and $z=3$ exhibit a very slight similarity to the target data at $z=0$.

We note that when provided with this progressively less informative (i.e., the density field at progressively earlier states), the network increasingly tends to default to outputting data that shows few dense structures (or even none) and is less correlated with the correct output, opting for more diffuse structures that can blend in with any target data's background, thus reducing on average the difference between prediction and any random target.

Given the previous observations, it is not surprising to find that the power spectrum (see Fig. \ref{fig:3D_TW_baseline_ps_ovl}, {\it left}) of the predicted data becomes lower for high $z$ inputs, since the lack of dense structures leads to overall lower density and a loss of signal at all frequencies. In terms of shape, we find that the predicted power spectra generally reproduce the overall shape of the target. When normalized, the spectra for $z=1$ and $z=0$ most closely follow the target shape, displaying a flatter plateau at intermediate $k$ followed by a sharper decline at high $k$, whereas predictions from $z=2$ and $z=3$ exhibit a smoother, more monotonous descent. The cross-correlation coefficient $r(k)$ remains high for $z=0$, with $r(k)\!\approx\!1$ up to $k\simeq10^{-1}\,h\,\mathrm{Mpc}^{-1}$, then gradually decreasing to zero at $k\simeq1\,h\,\mathrm{Mpc}^{-1}$. For $z=1$, we observed a typical linear-prediction behavior, with a good agreement at low $k$ but a much steeper drop beginning around $k\simeq4\times10^{-1}\,h\,\mathrm{Mpc}^{-1}$ and reaching $r(k)=0$ near $k\simeq2\times10^{-1}\,h\,\mathrm{Mpc}^{-1}$. Predictions from higher redshifts ($z=2,3$) decorrelate almost monotonically, from $r(k)\!\sim\!0.9$ at low $k$ to $r(k)=0$ at $k\simeq10^{-1}\,h\,\mathrm{Mpc}^{-1}$. These results confirm that the 3D case represents a more challenging task than the 2D configuration: the additional spatial dimension and greater variability in sub-cube mass and its evolution all compound the difficulty of reconstructing detailed structures, despite the potentially easier encoding of global features.

Finally, we examined the Dice coefficient (Fig. \ref{fig:3D_TW_baseline_ps_ovl}, {\it right}); once more the increased disparity between prediction and target with higher input $z$ is made clear, with the overall Dice value becoming significantly lower for input $z=1$ compared to $z=0$ and for inputs $z=2$ and $3$ compared to $z=1$. Overall, we can observe that the networks display a worse performance when applied to our 3D data compared to 2D.

\subsection{Introducing the velocity field}\label{vel_results}

\begin{figure}[h!]
    \includegraphics[width=.5\textwidth]{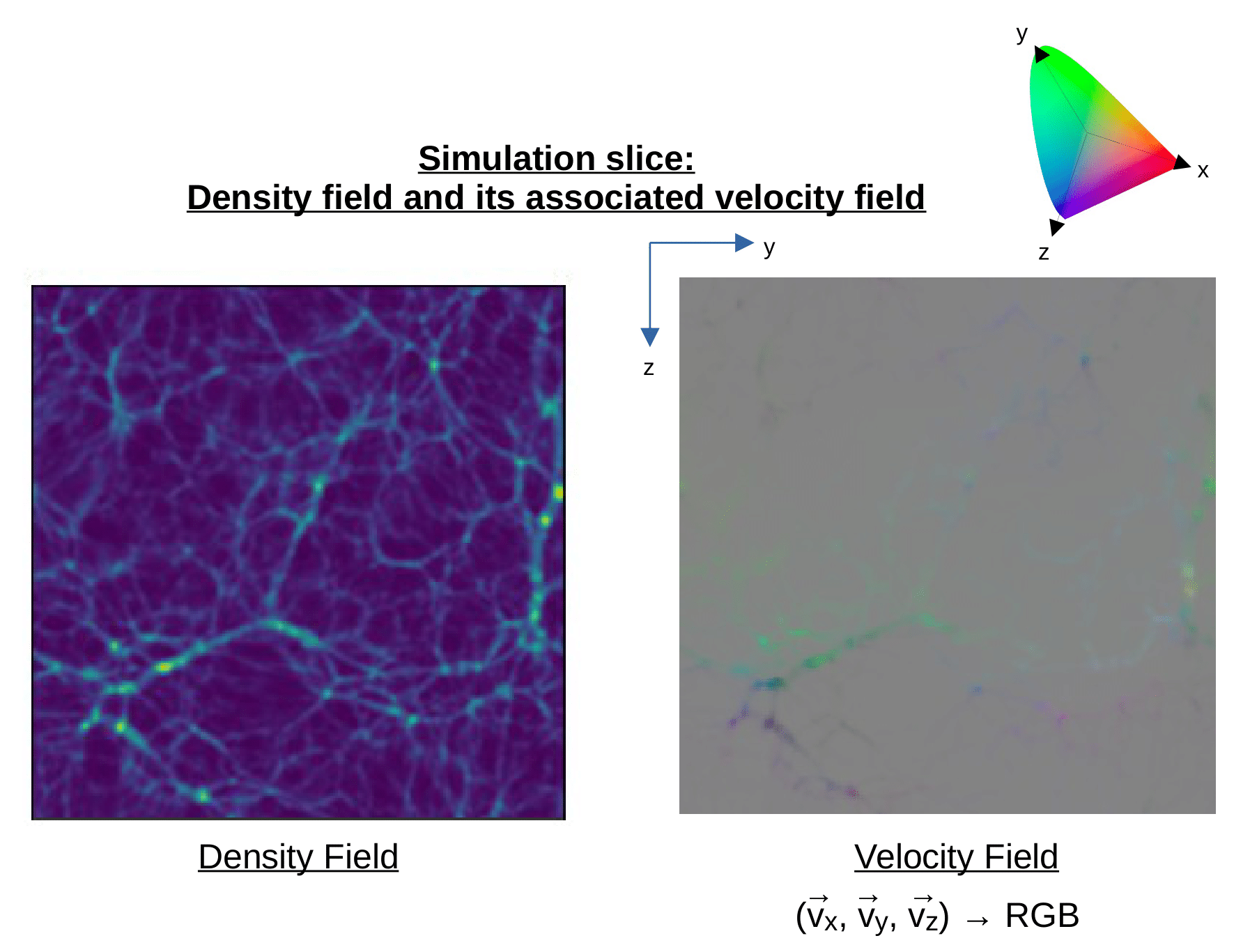}
    \caption{Example slice of a 3D simulation, showing the density field ({\it left}) and its associated velocity field ({\it right}), represented in ($\Vec{v}_x$,$\Vec{v}_y$,$\Vec{v}_z$) to (R,G,B)}
    \label{fig:vel_slice}
\end{figure}

Additional information in the form of input density fields' associated velocity fields (see Fig. \ref{fig:vel_slice}) is bound to constrain more thoroughly the target density fields, as initial velocities notably provide the necessary information for a linear evolution of the density field. These velocity fields can be obtained from the snapshots. Indeed, in their raw form and for every saved snapshot, our N-body simulations provide us with every particle's position, but also every particle's velocity, in the form of a 3D vector for each particle.

\begin{figure*}
    \centering
    \includegraphics[width=\textwidth]{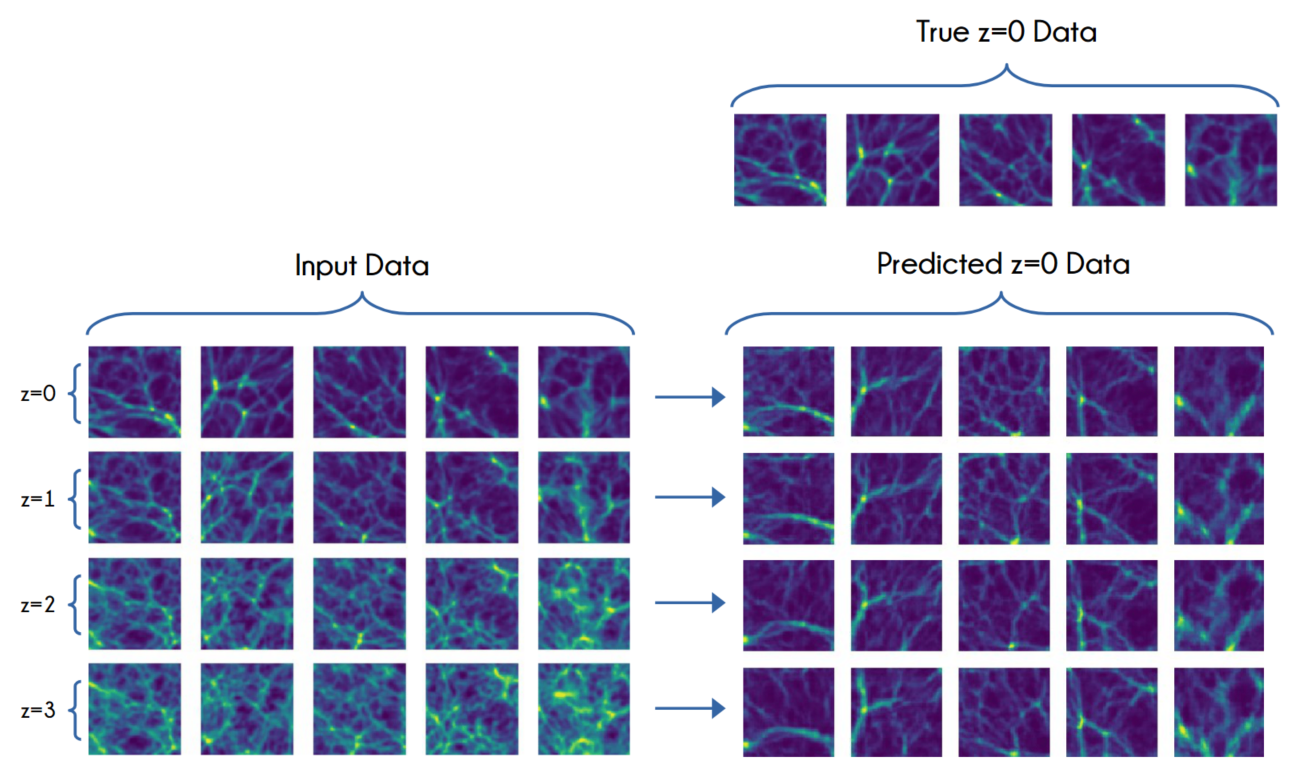}
    \caption{Five images from the 3D simulations at various redshifts ({\it left}), and their equivalent predictions of redshift $z=0$ ({\it right}) as inferred by the (density + velocity) TW. The true $z=0$ simulation images are shown above the predicted images ({\it upper right}) for comparison.}
    \label{fig:3D_TW_vel_ims}
\end{figure*}

Using these data, we trained a TW to recover the density field at $z=0$ from the input density + velocity fields ($\rho,v_x,v_y,v_z$) at redshifts varying from $z=0$ to $z=3$ by steps of $\delta z=1$. Observing the predicted data (Fig. \ref{fig:3D_TW_vel_ims}), we note that models yield nearly identical results regardless of the input $z$. Unsurprisingly, they seem to have the same limitations as the data inferred by the baseline TW with input $z=0$ (see Fig.\ref{fig:3D_TW_baseline_ims}), recovering a thicker dense structure more accurately than finer diffuse structure. Indeed, a TW trained with additional information should not exceed the results of an AE provided with all the required information as input.

\begin{figure*}
    \centering
    \includegraphics[width=\textwidth]{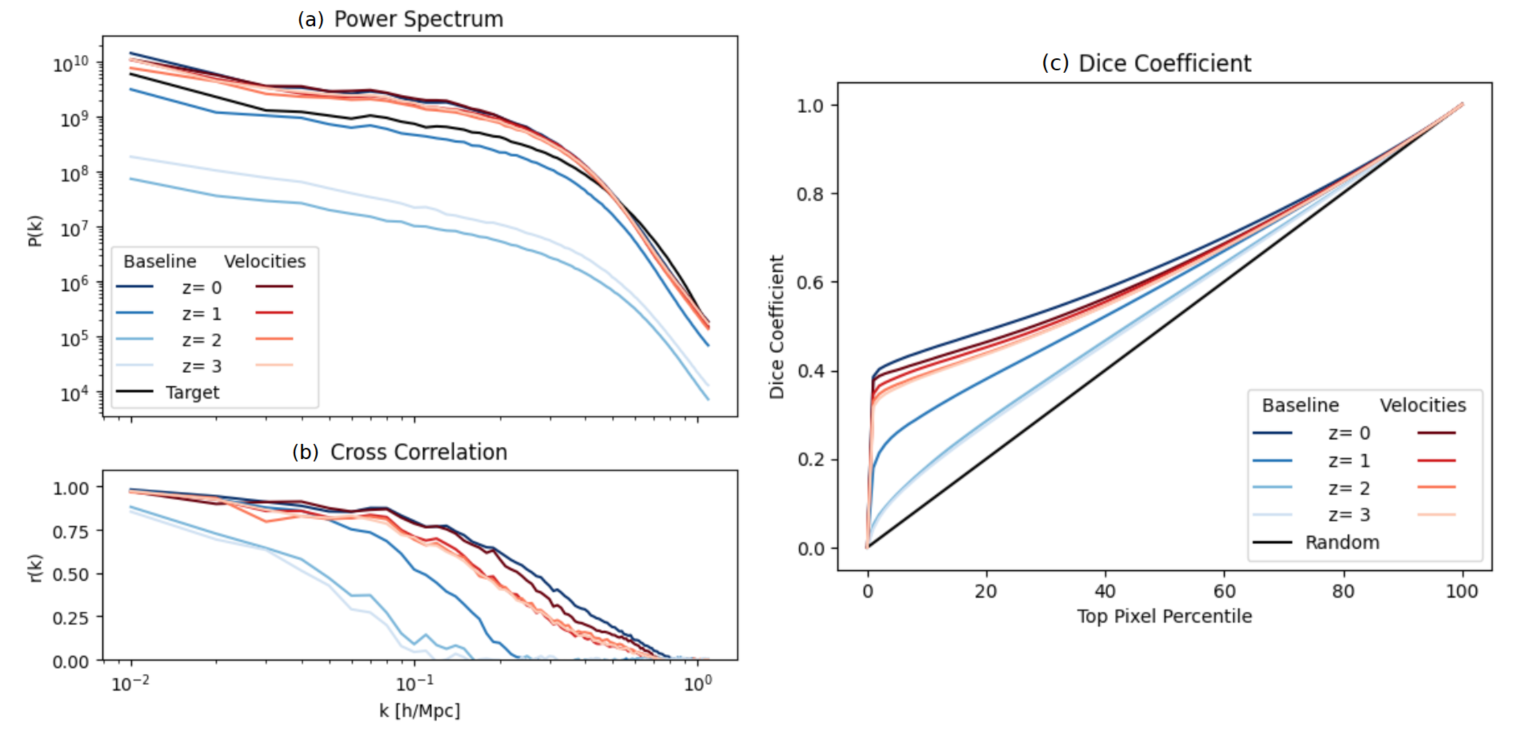}
    \caption{Spectra, cross-correlation and Dice coefficients for the {\it baseline} in blue, and {\it (density + velocity)} in red TW applied to 3D data.
{\it (a)} Power spectra from predictions from input redshifts z = 3 → 0 (blue/red scales), and target spectrum at z = 0 (black). 
{\it (b)} Corresponding cross-correlation coefficient between prediction and target over the same redshift range. {\it (c)} Dice coefficient between prediction and target (same color scales as in {\it a} and {\it b}).}
    \label{fig:3D_TW_vel_ps_ovl}
\end{figure*}

As can be expected given the visual similarity between target and predicted data, the predicted power spectra (Fig. \ref{fig:3D_TW_vel_ps_ovl}, {\it upper left}) are close, although slightly shifted upwards, to the true simulation power spectrum, especially when compared to those recovered by the {\it baseline TW}. Looking at the cross-correlation ({\it lower left}), we find that predictions from all input redshifts perform nearly as well as the simple $z=0\rightarrow0$ encoding case, with $r(k)\!\approx\!1$ up to $k\simeq10^{-1}\,h\,\mathrm{Mpc}^{-1}$, then gradually decreasing to zero at $k\simeq1\,h\,\mathrm{Mpc}^{-1}$. A study of the Dice coefficient (Fig. \ref{fig:3D_TW_vel_ps_ovl}, {\it right}) completes the picture by showing that the (density+velocity) TW outperforms the baseline TW, to the point that the recovered Dice of the (density+velocity) TW for input $z=3$ is better than that of the baseline TW with input $z=1$.

Finally, inspecting the evolution of the loss measured throughout training for both cases (baseline TW and velocities TW; i.e., density only vs. density+velocity input as seen in Fig. \ref{fig:3D_TW_vel_losses}), several noteworthy aspects emerge. First, past a certain point in training all training losses follow a clear linear trend in log-log. While significantly more noisy, this appears to be the case for validation losses as well. Comparing the two cases, we can note that density-only input training leads to overall greater and noisier loss, suggesting that adding velocity fields to the input makes the task less difficult and the training more stable.

We find that including the initial velocity field as input leads to significantly improved predictions of the evolved density field. This is not surprising: the velocity field encodes the direction and magnitude of matter flow, providing direct information about the dynamics driving structure formation. While the density field alone contains some of this information implicitly, supplying the velocity field explicitly allows the network to better anticipate structure evolution. This aligns with the broader physical understanding that matter trajectories (not just the initial positions) shape the development of cosmic structures.

\section{Discussion}\label{sec:discussion}

\subsection{Results and limitations}
The TW model was designed with a dual objective: to compress density fields into a compact representation with minimal information loss, and to predict their future evolution in redshift. This dual nature is central to understanding the performance of the model in different regimes.
In 2D, the baseline TW performs surprisingly well at the prediction task, with comparable performance metrics across various input redshifts. This suggests that the limiting factor in this configuration is primarily the encoding component; that is, once the data is well-encoded, predicting its evolution appears relatively easy. Improving the autoencoder structure (particularly its ability to faithfully reconstruct $z=0$ from itself) could therefore yield significant gains. Prior works on CAMELS \citep{villaescusa2021camels} have demonstrated that deep autoencoders have the capacity to encode this kind of data with high fidelity; albeit potentially at the cost of more complex and less semantically interpretable latent spaces.

Conversely, in 3D, the model exhibits significant degradation in prediction quality as the input redshift increases. This behavior indicates that in this case, the prediction task becomes the dominant bottleneck: as input redshift increases, forward mapping becomes increasingly difficult. We surmise that this is due to the higher-redshift density field inputs effectively providing less information about their $z=0$ descendants. 
While, in principle, a snapshot's full initial density field at the highest redshift ($z=19$) contains all the information required to predict the final state, this is due to our complete knowledge of the phase information (density field + initial velocity is 0) and of neighboring structures (periodic boundaries).

As time progresses, the density field by itself becomes an increasingly incomplete representation of the total phase-space information, although structures start to virialize and the large-scale distribution stabilizes. In our setup, the input subcubes do not possess periodic boundaries, which further limits the model’s access to the surrounding dynamical context. At intermediate and high redshifts ($z\gtrsim3$), neighboring regions can still significantly influence local evolution through matter inflows and outflows, which the model cannot infer from density alone. Thus, except perhaps at the very earliest times ($z\approx19$), higher redshift inputs effectively contain less predictive information about the $z=0$ configuration from the model’s limited perspective.

To ease this prediction challenge, we experimented with supplying the model with velocity fields in addition to density. This auxiliary information substantially improves performance; particularly by restoring nearly redshift-invariant prediction quality, as in the 2D case, leading once again to an encoding-based limitation.
Velocity information plays a crucial role in resolving ambiguities in the forward evolution of density fields by capturing both local matter dynamics and the influence of surrounding structures beyond the field of view. Thus, our results strongly suggest that the prediction limitations observed with density-only inputs stemmed primarily from insufficient input information, rather than from an inherent difficulty in approximating structure evolution over longer time intervals. 

\subsection{Theoretical interpretation}
We can reinterpret the TW’s operation as learning a complex, multiparameter function that maps an input density field to its future counterpart. More precisely, it maps from the space of density (and optionally velocity) fields to a latent representation and then to a constrained output space defined by a pretrained GAN's generative distribution.

This setup implicitly assumes two things; first, the existence and uniqueness of a ground truth: each input corresponds to a single expected output at $z=0$. Second, the inclusion of this output in the GAN’s generative space: the latent decoder must be able to represent the ground truth within its learned manifold.

The first assumption gets partially violated in practice. A single density field at a given redshift may evolve differently depending on information not present in the field itself, such as velocity vectors or neighboring structures. 
Thus, in our baseline 3D case, where only the density field is provided as input, the model must contend with a high variability in possible outcomes due to the limited information available. Faced with this uncertainty, the network adopts a conservative prediction strategy; rather than attempting to reconstruct stark high-density structures that could deviate strongly from the ground truth, it produces outputs with more diffuse structures, and less pronounced contrast. This behavior effectively minimizes the expected loss when the target is difficult to predict, but results in high-density structures being vastly underrepresented.

This explains the degradation in 3D predictions at higher redshifts, where forward evolution becomes increasingly uncertain, particularly for density-only inputs. The second assumption concerns the representational limits of the GAN. If the generative space cannot cover all plausible $z=0$ outcomes, the TW is inherently constrained to approximate the closest admissible output. This limitation is most evident when the decoder fails to reconstruct fine-scale structure, even in the "simple autoencoding" case where the input and target data are the same. Addressing this might require training more expressive generative models or exploring architectures that decouple the TW from such constraints entirely; although this risks compromising the realism of the output fields.

\section{Conclusion} \label{sec:conclusion}

This work presents a proof-of-concept approach for modeling the nonlinear evolution of large-scale structure using Eulerian fields as inputs. Unlike particle-based Lagrangian methods that excel at tracking individual trajectories, our model focuses on evolving continuous fields directly. This makes it a promising candidate for bridging toward applications that rely on field-level data, such as hydrodynamical simulations or observational reconstructions, where particle data is unavailable or ill-defined.
The architecture itself (i.e., a hybrid between an autoencoder and components drawn from GAN training) demonstrates that it is possible to balance local accuracy and statistical realism, although this often involves a tradeoff. In this first iteration, the model favors large-scale structure preservation and statistical agreement over precise pixel-level predictions. But this compromise is not fundamental: future models may learn to better balance these objectives or decouple them entirely.

Several avenues for future improvement can be considered, such as enhancing the model with stronger GAN variants or conditional GANs \citep{mirza2014conditional,antipov2017face}, which would open options such as adding cosmology dependance or training a single model for any input redshift, increasing latent space dimensionality (with potential costs to interpretability), and leveraging physical symmetries (i.e., isotropy or scale invariance) via architectures such as bispectral neural networks \citep{sanborn2022bispectral}. Further strategies might involve simplifying the prediction task through auxiliary fields (e.g., gravitational potential) or easing training with curriculum learning \citep{bengio2009curriculum, ullmo2022emulation}.
Beyond architectural improvements, alternate approaches such as VAE \citep{kingma2013auto}  or U-Nets \citep{siddique2021u} (as in the \citealt{he2019learning} article, but adapted for Eulerian fields) offer flexible modeling paradigms that are worth exploring or even combining with our method.
Finally, the model could be extended to tasks where statistical accuracy is more critical than exact reconstruction, such as denoising or inpainting masked regions. Another promising direction is inverse evolution: predicting earlier cosmic states from later ones:\ a challenging but potentially insightful task due to its intrinsic nonuniqueness \citep{Jasche_2019, Doeser_2024, legin2024posterior, jindal2023predicting}. Rather than aiming to replace the precision of Lagrangian simulators, this line of research seeks to complement them by offering efficient, adaptable tools for cosmological inference, particularly in cases where only field-level data is accessible.

\begin{acknowledgements}

The authors thank H. Tanimura for providing the 3D simulations, the IAS ByoPiC\footnote{\url{https://byopic.eu/}} and CRIStAL Sigma\footnote{\url{https://www.cristal.univ-lille.fr/equipes/sigma/}} teams for fruitful discussions and advice, and D. Jamieson for helpful information. We also thank the anonymous referee for their in-depth feedback, which greatly improved the quality of this manuscript. This project was funded by the European Research Council (ERC) under the European Union’s Horizon 2020 research and innovation programme grant agreement ERC-2015-AdG 695561. M.U. was supported by the Irfu of CEA Saclay, through the PTC program. A.D. was supported by the Comunidad de Madrid and the Complutense University of Madrid (Spain) through the Atraccián de Talento program (Ref. 2019-T1/TIC-13298).
\end{acknowledgements}

\bibliographystyle{aa}
\bibliography{main}

\end{document}